\documentclass[twocolumn,superscriptaddress,showpacs,prl]{revtex4}
\usepackage{graphicx,epsfig,psfrag}

\newcommand{\be}{\begin{equation}}
\newcommand{\ee}{\end{equation}}
\newcommand{\bea}{\begin{eqnarray}}
\newcommand{\eea}{\end{eqnarray}}
\newcommand{\al}{\alpha}

\newcommand{\lm}{\lambda}

\newcommand{\vep}{\varepsilon}

\newcommand{\nn}{\nonumber}
\newcommand{\grtsim}{\mbox{\raisebox{-3pt}{$\stackrel{>}{\sim}$}}}
\newcommand{\lessim}{\mbox{\raisebox{-3pt}{$\stackrel{<}{\sim}$}}}

\begin{document}

\title{
\boldmath 
Two-Loop Corrections to Bhabha Scattering
\unboldmath}
\author{Alexander A. Penin}
  \affiliation{Institut f\"ur Theoretische Teilchenphysik,
    Universit\"at Karlsruhe, 76128 Karlsruhe, Germany}
  \affiliation{Institute for Nuclear Research,
    Russian Academy of Sciences, 117312 Moscow, Russia}


\begin{abstract}
  The two-loop radiative photonic corrections to Bhabha scattering are
  computed in the leading order of the small electron mass expansion up
  to the nonlogarithmic term.  After including the soft photon
  bremsstrahlung we obtain the infrared-finite result for the
  differential cross section, which can directly be applied to a precise
  luminosity determination of the present and future $e^+e^-$ colliders.
\end{abstract}

\maketitle 

Electron-positron {\it Bhabha} scattering plays a special role in
particle phenomenology. It is crucial for extracting physics from
experiments at electron-positron colliders since it provides a very
efficient tool for luminosity determination.  The small angle Bhabha
scattering has been particularly effective as a luminosity monitor in
the LEP and SLC energy range because its cross section is large and QED
dominated \cite{Jad}.  At a future International Linear Collider the
luminosity spectrum is not monochromatic due to beam-beam effects.
Therefore measuring the cross section of the small angle Bhabha
scattering alone is not sufficient, and the angular distribution of the
large angle Bhabha scattering has been suggested for disentangling the
luminosity spectrum \cite{Too}.  The large angle Bhabha scattering is
important also at colliders operating at a center of mass energy
$\sqrt{s}$ of a few GeV, such as BABAR, BELLE, BEPC/BES, DA$\Phi$NE,
KEKB, PEP-II, and VEPP-2M, where it is used to measure the integrated
luminosity \cite{Car}. Since the accuracy of the theoretical evaluation
of the Bhabha cross section directly affects the luminosity
determination, remarkable efforts have been devoted to the study of the
radiative corrections to this process (see \cite{Jad} for an extensive
list of references).  Pure QED contributions are particularly important
because they dominate the radiative corrections to the large angle
scattering at intermediate energies 1-10~GeV and to the small angle
scattering also at higher energies.  The calculation of the QED
radiative corrections to the Bhabha cross section is among the classical
problems of perturbative quantum field theory with a long history.  The
first order corrections are well known (see \cite{Boh} and references
therein).  To match the impressive experimental accuracy the complete
second order QED effects have to be included on the theoretical side.
The evaluation of the two-loop virtual corrections constitutes the main
problem of the second order analysis.  The complete two-loop virtual
corrections to the scattering amplitudes in the massless electron
approximation have been computed in Ref.~\cite{BDG}, where dimensional
regularization has been used for infrared divergences.  However, this
approximation is not sufficient since one has to keep a nonvanishing
electron mass to make the result compatible with available Monte Carlo
event generators \cite{Jad}.  Recently an important class of the
two-loop corrections, which include at least one closed fermion loop,
has been obtained for a finite electron mass \cite{BFMR} including the
soft photon bremsstrahlung \cite{Bon}.  A similar evaluation of the
purely photonic two-loop corrections is a challenging problem at the
limit of present computational techniques \cite{HeiSmi}.  On the other
hand in the energy range under consideration only the leading
contribution in the small ratio $m_e^2/s$ is of phenomenological
relevance and should be retained in the theoretical estimates. In this
approximation all the two-loop corrections enhanced by a power of the
large logarithm $\ln(m_e^2/s)$ are known so far for the small angle
\cite{Arb} and large angle \cite{AKS,GTB} Bhabha scattering while the
nonlogarithmic contribution is still missing.

In this Letter we complete the calculation of the two-loop radiative
corrections in the leading order of the small electron mass expansion.
For this purpose we develop the method  of {\it infrared
  subtractions}, which simplifies the calculation by fully exploiting
the information on the general structure of infrared singularities in
QED.

The leading asymptotics of the virtual corrections cannot be obtained
simply by putting $m_e=0$ because the electron mass regulates the
collinear divergences. In addition the virtual corrections are a subject
of soft divergences, which can be regulated by giving the photon a small
auxiliary mass $\lm$. The soft divergences are canceled out in the
inclusive cross section when one adds the contribution of the soft
photon bremsstrahlung \cite{KLN}.  Here we should note that the
collinear divergences in the massless approximation are also canceled in
a cross section which is inclusive with respect to real photons
collinear to the initial or final state fermions \cite{SteWei}. This
means that if an angular cut on the collinear emission is sufficiently
large, $\theta_{cut}\gg \sqrt{m_e^2/s}$, the inclusive cross section is
insensitive to the electron mass and can in principle be computed with
$m_e=0$ by using dimensional regularization of the infrared divergences
for both virtual and real radiative corrections like it is done in the
theory of QCD jets.  However, as it has been mentioned above, all the
available Monte Carlo event generators for Bhabha scattering with
specific cuts on the photon bremsstrahlung dictated by the experimental
setup employ a nonzero electron mass as an infrared regulator, which
therefore has to be used also in the calculation of the virtual
corrections. Thus we have to compute the two-loop virtual corrections to
the four-fermion amplitude ${\cal A}^{(2)}(m_e,\lm)$.  The general
problem of the calculation of the small mass asymptotics of the
corrections including the power-suppressed terms can systematically be
solved within the expansion by regions approach \cite{Smi}. To get the
leading term in $m_e^2/s$ we develop the method applied first in
Ref.~\cite{FKPS} to the analysis of the two-loop corrections to the
fermion form factor in an Abelian gauge model with mass gap, which is
briefly outlined below.  The main idea is to construct an auxiliary
amplitude $\tilde{\cal A}^{(2)}(m_e,\lm)$, which has the same structure
of the infrared singularities but is simpler to evaluate.  Then the
difference ${\cal A}^{(2)}-\tilde {\cal A}^{(2)}$ has a finite limit
$\delta{\cal A}^{(2)}$ as $m_e,~\lm\to 0$. This quantity does not depend
on the regularization scheme for ${\cal A}^{(2)}$ and $\tilde{\cal
  A}^{(2)}$ and can be evaluated in dimensional regularization in the
limit of four space-time dimensions.  In this way we obtain ${\cal
  A}^{(2)}(m_e,\lm)= \tilde {\cal A}^{(2)}(m_e,\lm)+\delta{\cal
  A}^{(2)}+{\cal O}(m_e,\lm)$.  The singular dependence of the virtual
corrections on infrared regulators obeys evolution equations, which
imply factorization of the infrared singularities \cite{Mue}. One can
use this property to construct the auxiliary amplitude $\tilde{\cal
  A}^{(2)}(m_e,\lm)$.  For example, the collinear divergences are known
to factorize into the external line corrections \cite{Fre}. This means
that the singular dependence of the corrections to the four-fermion
amplitude on $m_e$ is the same as of the corrections to (the square of)
the electromagnetic fermion form factor.  The remaining singular
dependence of the amplitude on $\lm$ satisfies a linear differential
equation \cite{Mue} and the corresponding soft divergences exponentiate.
A careful analysis shows that for pure photonic corrections $\tilde{\cal
  A}^{(2)}(m_e,\lm)$ can be constructed of the two-loop corrections to
the form factor and the products of the one-loop contributions. Note
that the corrections have matrix structure in the chiral amplitude basis
\cite{KMPS}. We have checked that in dimensional regularization the
structure of the infrared divergences of the auxiliary amplitude
obtained in this way agrees with the one given in Refs.~\cite{BDG,Cat}.
Thus in our method the infrared divergences, which induce the asymptotic
dependence of the virtual corrections on the electron and photon masses,
are absorbed into the auxiliary factorized amplitude while the
technically most nontrivial calculation of the matching term
$\delta{\cal A}^{(2)}$ is performed in the massless approximation.  Note
that the method does not require a loop-by-loop subtracting of the
infrared divergences since only a general information on the infrared
structure of the total two-loop correction is necessary to construct
$\tilde{\cal A}^{(2)}(m_e,\lm)$.  Clearly, the method can be adopted to
different amplitudes, mass spectra and number of loops.  For the
calculation of the matching term $\delta{\cal A}^{(2)}$ beside the
one-loop result one needs the two-loop corrections to the four-fermion
amplitude and to the form factor in massless approximation, which are
available \cite{BDG,KraLam}.  For the calculation of $\tilde{\cal
  A}^{(2)}(m_e,\lm)$ one needs also the two-loop correction to the form
factor for $\lm\ll m_e\ll s$ which can be found in Ref.~\cite{MasRem} as
a specific limit of the result for an arbitrary momentum transfer.  We
independently cross-checked it by integrating the dispersion relation
with the spectral density computed in Ref.~\cite{BMR}.  The closed
fermion loop contribution, which is included into the analysis
\cite{MasRem,BMR}, can be separated by using the result of
Ref.~\cite{Bur}.  Let us now present our result.  We define the
perturbative expansion for the Bhabha cross section in the fine
structure constant $\al$ as follows:
$\sigma=\sum_{n=0}^\infty\left({\al\over\pi}\right)^n\sigma^{(n)}$.  The
leading order differential cross section reads
\be
{{\rm d}\sigma^{(0)}\over{\rm d}\Omega}
={\al^2\over s}\left({1-x+x^2\over x}\right)^2\,,
\label{losig}
\ee 
where $x=(1-\cos\theta)/2$ and $\theta$ is the scattering angle.
Virtual corrections taken separately are infrared divergent.  To get a
finite scheme independent result we include the contribution of the soft
photon bremsstrahlung into the cross section.  Thus the second order
corrections can be represented as a sum of three terms
$\sigma^{(2)}=\sigma^{(2)}_{vv}+\sigma^{(2)}_{sv}+\sigma^{(2)}_{ss}$,
which correspond to the two-loop virtual correction including the
interference of the one-loop corrections to the amplitude, one-loop
virtual correction to the single soft photon emission, and the double
soft photon emission, respectively. When the soft photon energy cut is
much less than $m_e$, the result for the two last terms in the above
equation is known in analytical form and can be found {\it e.g.} in
Ref.~\cite{AKS}.  The second order correction can be decomposed 
according to the asymptotic dependence on the electron mass
\be
{{\rm d}\sigma^{(2)}\over{\rm d}\sigma^{(0)}}=
\delta^{(2)}_{2}\ln^2\left({s\over m_e^2}\right)
+\delta^{(2)}_{1}\ln\left({s\over m_e^2}\right)
+\delta^{(2)}_{0}+{\cal O}(m_e^2/s)\,,
\label{decom}
\ee 
where the coefficients $\delta_i^{(j)}$ are independent on $m_e$.  The
result for the logarithmically enhanced corrections is known (see
\cite{AKS} and references therein). For the nonlogarithmic term we
obtain
\begin{widetext}
\bea
\lefteqn{\delta^{(2)}_{0}=
8{\cal L}_{\vep}^2
+\left({1-x+x^2}\right)^{-2}\bigg[\left({4\over 3}
-{8\over 3}x-x^2+{10\over 3}x^3-{8\over 3}x^4\right)\pi^2
+\left(-12+16x-18x^2+6x^3\right)\ln(x)
\nn }\\
&&
+\left(2x+2x^3\right)\ln(1-x)+\left(-3x+x^2+3x^3-4x^4\right)
\ln^2(x)+\left(-8+16x-14x^2+4x^3\right)\ln(x)\ln(1-x)
\nn \\
&&
+\left(4-10x+14x^2-10x^3+4x^4\right)\ln^2(1-x)
+\left({1-x+x^2}\right)^2(16+8{\rm Li_2}(x)-8{\rm Li_2}(1-x))
\bigg]{\cal L}_{\vep}+\left({1-x+x^2}\right)^{-2}
\nn \\
&&
\times\Bigg(\!\left(1 - x + x^2\right)^2\left(8+4f_0^{(2)}-2{f_0^{(1)}}^2\right)
+\left({4\over 3}-{23\over 24}x-{25\over 8}x^2+{121\over 24}x^3
-{19\over 6}x^4\right)\pi^2+\left({1\over 18} - {203\over 360}x + 
{25\over 36}x^2-{41\over 180}x^3\right.
\nn \\
&&
\left. + 
{109\over 1440}x^4\right)\pi^4
+\left({7\over 2}x-7x^2+4x^3\right)\zeta(3)
+\left[-{93\over 8}+{231\over 16}x-{279\over 16}x^2+{93\over 16}x^3
\right.+\left(-{3\over 2}+{13\over 4}x-{7\over 12}x^2
-{11\over 8}x^3\right)\pi^2
\nn \\
&&
+\left(12-12x+8x^2-x^3\right)\zeta(3)\bigg]\ln(x)
+\left[{9\over 2}-{43\over 8}x+{17\over 8}x^2
+{29\over 8}x^3-{9\over 2}x^4+\left({x\over 4}+{x^2\over 2}+{5\over 24}x^3
+{19\over 48}x^4\right)
\pi^2\right]\ln^2(x)
\nn \\
&&
+\left({67\over 24}x-{5\over 4}x^2-{2\over 3}x^3\right)\ln^3(x)
+\left({7\over 48}x+{5\over 96}x^2-{x^3\over 12}
+{43\over 96}x^4\right)\ln^4(x)+\bigg\{3x+3x^3+\left({7\over 6}x-{73\over 24}x^2
+{15\over 8}x^3\right)\pi^2
\nn \\
&&
+\left(-6+6x-x^2-4x^3\right)\zeta(3)
+\left[-8+{21\over 2}x-{45\over 4}x^2+x^4 
+\left(1-{x\over 6}+{x^2\over 12}
-{x^3\over 3}
-{x^4\over 8}\right)\pi^2\right]\ln(x)+\bigg(6-11x
\nn \\
&&
\left.+{35\over 4}x^2-{15\over 8}x^3\right)\ln^2(x)
+\left({2\over 3}+{x\over 12}-{x^3\over 3}
+{5\over 24}x^4\right)\ln^3(x)\bigg\}\ln(1-x)
+\bigg[{7\over 2}-6x+{45\over 4}x^2-6x^3+{7\over 2}x^4
+\left(-{17\over 24}\right.
\nn \\
&&
\left.+{7\over 6}x-{25\over 24}x^2
-{13\over 48}x^4\right)\pi^2
+\left(-3+{23\over 4}x-{23\over 4}x^2
+{9\over 8}x^3\right)\ln(x)
+\left({7\over 2}-{41\over 8}x+{31\over 8}x^2+{3\over 8}x^3
-{13\over 16}x^4\right)\ln^2(x)\bigg]
\nn \\
&&
\times\ln^2(1-x)+\left[{3\over 8}x+{1\over 6}x^2+{3\over 8}x^3
+\left(-4+{29\over 6}x-{49\over 12}x^2+{5\over 6}x^3
+{7\over 8}x^4\right)\ln(x)\right]\ln^3(1-x)
+\left({1\over 32}-{3\over 4}x+{71\over 48}x^2\right.
\nn \\
&&
\left.
-{29\over 24}x^3+{9\over 32}x^4\right)
\ln^4\left(1-x\right)
+\bigg\{8-16x+24x^2-16x^3+8x^4
+\left({7\over 3}-3x+{3\over 4}x^2
+{5\over 6}x^3-{2\over 3}x^4\right)\pi^2
+\left[-6+{11\over 2}x\right.
\nn \\
&&
\left.-4x^2+x^3+\left(2
-{11\over 4}x+{7\over 4}x^2
+{x^3\over 4}-x^4\right)\ln(x)\right]\ln(x)
+\left[{3\over 2}x-{x^2\over 4}+x^3
+\bigg(-4+9x-{15\over 2}x^2+2x^3\right)\ln(x)
\nn \\
&&
+\left(-1-{7\over 2}x+{25 \over 4}x^2
-5x^3+2x^4\right)\ln(1-x)\bigg]\ln(1-x)
+\left(2-4x+6x^2-4x^3+2x^4\right){\rm Li}_2(x)\bigg\}{\rm Li}_2\left(x\right)
+\bigg\{-8
\nn \\
&&
+16x-24x^2+16x^3-8x^4 
+\left[-{2\over 3}+{4\over 3}x+{x^2\over 2}-{5\over 3}x^3
+{2\over 3}x^4\right]\pi^2
+\bigg[6-8x+9x^2-3x^3+\left({3\over 2}x-{x^2\over 2}
-{3\over 2}x^3\right.
\nn \\
&&
\left.
+2x^4\bigg)\ln(x)\right]\ln(x)
+\left[-x-{x^2\over 4}-{x^3\over 2}+\left(10-14x+9x^2\right)\ln(x)
+\left(-8+11x-{31\over 4}x^2+{x^3\over 2}
+x^4\right)\ln(1-x)\right]
\nn \\
&&
\times\ln(1-x)
+\left(-4+8x-12x^2+8x^3
-4x^4\right){\rm Li}_2(x)
+\left(2-4x+6x^2-4x^3+2x^4\right){\rm Li}_2(1-x)\bigg\}
{\rm Li}_2\left(1-x\right)+
\left[{5\over 2}x\right.
\nn \\
&&
\left.-5x^2+2x^3+\left(-4-x+x^2+2x^3
-2x^4\right)\ln(x)+\left(6-6x+x^2+4x^3\right)\ln(1-x)\right]
{\rm Li}_3\left(x\right)+\left[{x\over 2}-{x^3\over 2}
+\big(-6+5x\right.
\nn \\
&&
\left.\left.+3x^2-5x^3\right)\ln(x)
+\left(6-10x+10x^3
-6x^4\right)\ln(1-x)\right]
{\rm Li}_3\left(1-x\right)
+\left(-2+{17\over 2}x-{17\over 2}x^3
+2x^4\right){\rm Li}_4\left(x\right)
\nn\\
&&
+\left(7x-{9\over 2}x^2-4x^3+6x^4\right){\rm Li}_4\left(1-x\right)
+\left(-6+4x+{9\over 2}x^2-7x^3\right){\rm Li}_4\left(-{x\over 1-x}\right)
\Bigg)\,,
\label{result}
\eea
\end{widetext}
where $\zeta(3)=1.202057\ldots$ is the value of the Riemann's
zeta-function, ${\rm Li}_n(z)$ is the polylogarithm, 
${\cal  L}_{\vep}=\left[1-\ln\left(x/(1-x)\right)\right]
\ln\left(\vep_{cut}/\vep\right)$, $\vep=\sqrt{s}/2$, 
and $\vep_{cut}$ is the energy cut on each emitted soft photon. In
Eq.(\ref{result}) $f_0^{(n)}$ stands for the $n$-loop nonlogarithmic
coefficient in the series for the form factor at Euclidean momentum
transfer
\bea
f_0^{(1)}&=&\pi^2/12-1\,,
\nn\\
f_0^{(2)}&=&{15\over 8}+{43\over 96}\pi^2-{59\over 1440}\pi^4
-{\pi^2\ln 2\over 2} - {9\over 4}\zeta(3)\,.
\label{formf}
\eea
As it follows from the generalized eikonal representation \cite{Fad},
in the limit of small scattering angles the two-loop corrections to
the cross section are completely determined by the corrections to the
electron and positron form factors in the $t$-channel amplitude.  In
the limit $x\to 0$ Eq.~(\ref{result}) agrees with the
asymptotic small angle expression given in \cite{Arb,Fad}, which is
a quite nontrivial check of our result. Note  that 
Eq.~(\ref{result}) is not valid  for very small scattering angles
corresponding to  $x\lessim m_e^2/s$ and for almost backward scattering
corresponding to $1-x\lessim m_e^2/s$, where the power-suppressed terms
become important.
 
\begin{figure}[t]
\epsfig{figure=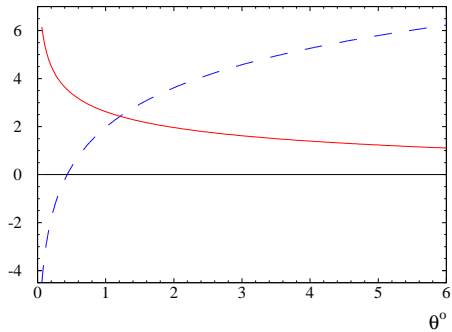,height=5cm}
\caption{\label{fig1} 
  Logarithmically enhanced (dashed line) and nonlogarithmic (solid line)
  second order corrections to the differential cross section of the
  small angle Bhabha scattering as functions of the scattering angle for
  $\sqrt{s}=100$~GeV and $\ln(\vep_{cut}/\vep)=0$, in units of
  $10^{-3}$. }
\end{figure}

\begin{figure}[t]
\epsfig{figure=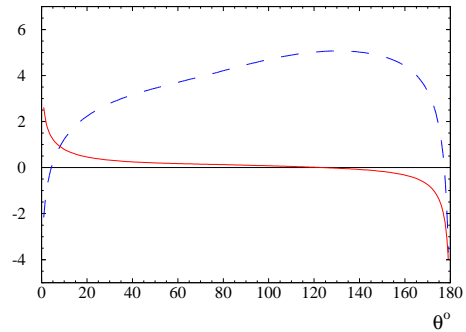,height=5cm}
\caption{\label{fig2} 
  The same as Fig.~(\ref{fig1}) but for the large angle Bhabha
  scattering and  $\sqrt{s}=1$~GeV.  }
\end{figure}

The second order corrections to the differential cross section
$(\al/\pi)^2{\rm d}\sigma^{(2)}/{\rm d}\sigma^{(0)}$ are plotted as
functions of the scattering angle for the small angle Bhabha scattering
at $\sqrt{s}=100$~GeV on Fig.~(\ref{fig1}) and for the large angle Bhabha
scattering at $\sqrt{s}=1$~GeV on Fig.~(\ref{fig2}).  We separate the
logarithmically enhanced corrections given by the first two terms of
Eq.~(\ref{decom}) and nonlogarithmic contribution given by the last term
of this equation.  All the terms involving a power of the logarithm
$\ln(\vep_{cut}/\vep)$ are excluded from the numerical estimates because
the corresponding contribution critically depends on the event selection
algorithm and cannot be unambiguously estimated without imposing
specific cuts on the photon bremsstrahlung.  We observe that for
scattering angles $\theta \lessim 18^o$ and $\theta \grtsim 166^o$ the
nonlogarithmic contribution exceeds $0.05\%$, and in the low energy
case exceeds the logarithmically enhanced contribution for $\theta
\lessim 8^o$.

To conclude, we have derived the two-loop radiative photonic corrections
to  Bhabha scattering in the leading order of the small electron mass
expansion up to nonlogarithmic term.  The nonlogarithmic contribution
has been found numerically important for the practically interesting
range of scattering angles.  Together with the result of
Refs.~\cite{BFMR,Bon} for the corrections with the closed fermion loop
insertions our result gives a complete expression for the two-loop
virtual corrections. It should be incorporated into the Monte
Carlo event generators to reduce the theoretical error in the luminosity
determination at present and future electron-positron colliders below
$0.1\%$.

\begin{acknowledgments}
  We would like to thank V.A. Smirnov for his help in the evaluation of
  Eq.(\ref{formf}) and O.L. Veretin for pointing out that the method of
  Ref.~\cite{FKPS} can be applied to the analysis of Bhabha scattering.
  We are grateful to R. Bonciani and A. Ferroglia \cite{BonFer} for cross-checking a
  large part of the result.  We thank J.H. K\"uhn and M. Steinhauser for
  carefully reading the manuscript and useful comments. The work was
  supported in part by BMBF Grant No.\ 05HT4VKA/3 and
  Sonderforschungsbereich Transregio 9.
\end{acknowledgments}

\end{document}